\documentstyle[preprint,eqsecnum,aps,floats,epsf]{revtex}

\def\beq{\begin{eqnarray}}
\def\eeq{\end{eqnarray}}

\def\non{\nonumber}
\def\la{\langle}
\def\ra{\rangle}
\def\ep{\varepsilon}

\def\Ova{O^q_{V-A}}
\def\Osp{O^q_{S-P}}
\def\Tva{T^q_{V-A}}
\def\Tsp{T^q_{S-P}}

\def\lsim{ {\ \lower-1.2pt\vbox{\hbox{\rlap{$<$}\lower5pt\vbox{\hbox{$\sim$}
}}}\ } }

\begin{document}
\preprint{
\font\fortssbx=cmssbx10 scaled \magstep2
\hbox to \hsize{
\hfill$\raise .5cm\vtop{
                \hbox{IP-ASTP-03-98}}$}
}
\draft
\vfill
\title{ Nonspectator Effects and $B$ Meson Lifetimes\\
from a Field-theoretic  Calculation}
\draft
\vfill
\author{Hai-Yang Cheng \footnote{Email address:
{\tt PHCHENG@ccvax.sinica.edu.tw}} and
Kwei-Chou Yang \footnote{Email address: {\tt 
kcyang@phys.sinica.edu.tw}}} 
\address{Institute of Physics, Academia Sinica, Taipei, Taiwan 115, R.O.C.}

\maketitle
\date{June 1997}
\begin{abstract}
The $B$ meson lifetime ratios are calculated to the order of
$1/m_b^3$ in the heavy quark expansion. The predictions of those
ratios are dependent on four unknown hadronic parameters $B_1$, $B_2$, 
$\varepsilon_1$ and $\varepsilon_2$, where $B_1$ and $B_2$ parametrize 
the matrix elements of color singlet-singlet
four-quark operators and $\varepsilon_1$ and $\varepsilon_2$ the matrix 
elements of color octet-octet operators. We derive the
renormalization-group improved QCD sum rules for these
parameters within the framework of heavy quark effective theory.
The results are $B_1(m_b)=0.96\pm 0.04$, $B_2(m_b)=0.95\pm 0.02$,
$\varepsilon_1(m_b)=-0.14\pm 0.01$, and $\varepsilon_2(m_b)=-0.08\pm 0.01$
to zeroth order in $1/m_b$. The resultant $B$ meson lifetime ratios 
are $\tau(B^-)/\tau(B_d)=1.11\pm 0.02$ and
$\tau(B_s)/\tau(B_d)\approx 1$ in SU(3) symmetry limit.
\end{abstract}

\vspace{0.7in}
\section{Introduction}

A QCD-based formulation for treatment of inclusive heavy hadron decays
has been developed in past years \cite{BIGI,Blok,Manohar}. According to 
the optical theorem, 
the inclusive decay rates are related to the imaginary part of certain
forward scattering amplitudes along the physical cut. Since the
cut is dominated by physical intermediate hadron states like resonances
which are nonperturbative in nature, {\it a priori} the operator
product expansion (OPE) or heavy quark expansion cannot be carried out
on the physical cut.
Nevertheless, for inclusive semileptonic decays, OPE can be employed for some
smeared or averaged physical quantities.
For example, by integrating out the neutrino energy, one can apply the OPE
to the double differential cross section $d^2\Gamma/(dq^2dE_\ell)$ by 
deforming the contour of integration into the unphysical region far away
from the physical cut \cite{Chay}.
Therefore, global quark-hadron duality, namely the matching
between the hadronic and OPE-based expressions for decay widths and
smeared spectra in semileptonic $B$ or bottom baryon decays, follows from the 
OPE and is justified 
except for a small portion of the contour near the physical cut which is
of order $\Lambda_{\rm QCD}/m_Q$.
Unfortunately, there is no analogous
variable to be integrated out in inclusive nonleptonic decays,
allowing an analytic continuation into
the complex plane. As a result, one has to invoke the assumption of 
local quark-hadron duality in order to appply the OPE in the physical region 
\cite{Falk}.
It is obvious that local duality is theoretically less firm and
secure than global duality. In order to test the validity of local 
quark-hadron duality, it is thus very important to have a reliable
estimate of the heavy hadron lifetimes within the OPE framework and
compare them with experiment.
 
  In the heavy quark limit, all bottom hadrons have the same lifetimes,
a well-known result in the parton picture. With the advent of 
heavy quark effective theory and the
OPE approach for the analysis of inclusive weak decays, it is realized
that the first nonperturbative correction to bottom hadron lifetimes 
starts at order $1/m_b^2$
and it is model independent (for a review, see \cite{Big}).
However, the $1/m_b^2$ corrections are small
and essentially canceled out in the lifetime ratios. 
The nonspectator effects such as $W$-exchange and
Pauli interference due to four-quark interactions are of order $1/m_Q^3$, 
but their contributions can be potentially significant due to a 
phase-space enhancement
by a factor of $16\pi^2$. As a result, the
lifetime differences of heavy hadrons come mainly from the above-mentioned 
nonspectator effects.

The world average values for the lifetime ratios 
of bottom hadrons are \cite{LEP}:
\begin{eqnarray}\label{taudata}
   {\tau(B^-)\over\tau(B^0_d)} &=& 1.07\pm 0.04 \,, \nonumber\\
   {\tau(B^0_s)\over\tau(B^0_d)} &=& 0.95\pm 0.05 \,, \nonumber\\
\frac{\tau(\Lambda_b)}{\tau(B^0_d)} &=& 0.78\pm 0.06 \,.
\end{eqnarray}
Since the model-independent prediction of $\tau(\Lambda_b)/\tau(B_d)$
to order $1/m_b^2$ is very close to unity [see Eq.~(\ref{taucrude}) below],
the conflict between theory and experiment for this lifetime ratio is
quite striking and has received a lot of attention 
\cite{Uraltsev,Altarelli,Col,NS,Cheng,Lipkin,Ito,BLLS}. 
One possible reason for the discrepancy
is that local quark-hadron duality may not work in the study of nonleptonic 
inclusive decay widths. Another possibility is that some hadronic matrix 
elements of four-quark operators are probably
larger than what naively expected so that the nonspectator effects of order
$16\pi^2/m_b^3$ may be large enough to explain the observed lifetime 
difference between the $\Lambda_b$ and $B_d$.
Therefore, as stressed by Neubert and Sachrajda~\cite{NS},
one cannot conclude that local duality truly fails before a reliable
field-theoretical calculation of the four-quark matrix elements is obtained.
Contrary to the $1/m_b^2$ corrections, the estimate of the nonspectator 
effects is, unfortunately, quite model dependent.
 
Conventionally, the hadronic matrix elements of four-quark operators are
evaluated using the factorization approximation for mesons and the quark
model for baryons. However, as we shall see, nonfactorizable 
effects absent in the factorization hypothesis can affect the $B$ meson 
lifetime ratios significantly. In order to
have a reliable estimate of the hadronic 
parameters $B_1$, $B_2,~\varepsilon_1$
and $\varepsilon_2$ in the meson sector, to be introduced below,
we will apply the QCD sum rule to calculate these unknown parameters.
After a brief review on the
status of the OPE approach for the $B$ hadron lifetime ratios in Sec.~II,
we proceed to derive in Sec.~III the
renormalization-group improved QCD sum rules for the parameters
$B_i$ and $\varepsilon_i$ and present a detailed analysis.
Sec. IV gives discussions and conclusions.

\section{A brief overview}
Within  the heavy quark expansion framework, we will focus in this paper 
on the study of the four-quark matrix elements of the $B$ meson
to understand the problem with $B$ meson lifetime ratios.
Before proceeding, let us briefly review the content of the theory.
Applying the optical theorem, the inclusive decay width of the bottom hadron
$H_b$ containing a $b$ quark can
be expressed in the form
\begin{equation}\label{imt}
   \Gamma(H_b\to X) = \frac{1}{m_{H_b}}\,\mbox{Im}\,
  \!\int{\rm d}^4x\, \langle H_b|\,T\{\,i
   {\cal L}_{\rm eff}(x),{\cal L}_{\rm eff}(0)\,\}
\,|H_b\rangle \,,
\end{equation}
where ${\cal L}_{\rm eff}$ is the relevant effective weak Lagrangian
that contributes to the particular final state $X$.
When the energy release in a $b$ quark decay is sufficiently large,
it is possible to express the nonlocal operator product in Eq.~(\ref{imt})
as a series of local operators in powers  of $1/ m_b$
by using the OPE technique.
In the OPE series, the only locally gauge invariant operator with
dimension four, $\bar b i\!\! \not\!\! D b$, can be reduced to $m_b \bar bb$
by using the equation of motion. Therefore, the first nonperturbative 
correction to the
inclusive $B$ hadron decay width starts at order $1/m_b^2$.\footnote{It is
emphasized in \cite{Akhoury} that the cancellation of the $1/m_Q$ corrections 
to the inclusive decay width occurs when it is expressed in terms of the 
running short-distance quark mass, e.g. the $\overline{\rm MS}$ mass, rather 
than the pole quark mass.}
As a result, the inclusive decay width of a hadron $H_b$ can
be expressed as~\cite{BIGI,Blok}
\begin{eqnarray}\label{gener}
   \Gamma(H_b\to X) =&& \frac{G_F^2 m_b^5 |V_{\rm CKM}|^2}{192\pi^3}\,
\frac{1}{2m_{H_b}}
   \left\{  c_3^X\,\langle H_b|\bar b b|H_b\rangle
+ c_5^X\,
\frac{\langle H_b|\bar b\,{1\over 2} g_s\sigma_{\mu\nu} G^{\mu\nu} b 
|H_b\rangle} {m_b^2}\right. \nonumber\\
    && \left.+ \sum_n c_6^{X(n)}\,\frac{\langle H_b| O_6^{(n)}|H_b 
    \rangle}{m_b^3} +  O(1/m_b^4) \right\} \,,
\end{eqnarray}
where $V_{\rm CKM}$ denotes some combination of the Cabibbo-Kobayashi-Maskawa 
parameters and $c_i^X$ reflect short-distance dynamics and phase-space
corrections.
The matrix elements in Eq.~(\ref{gener}) can be systematically 
expanded in powers of $1/m_b$ in heavy quark effective theory 
(HQET)~\cite{review}, in which the $b$-quark field is represented by a
four-velocity-dependent field denoted by $h^{(b)}_v(x)$. 

In Eq.~(\ref{gener}) $c^X_i$ are functions of $c_1$ and $c_2$, the 
Wilson coefficients in the effective Hamiltonian 
\beq
{\cal H}^{\Delta B=1}_{\rm eff} &=& {G_F\over\sqrt{2}}\Big[ V_{cb}V_{uq}^*(c_1
(\mu)O_1^u(\mu)+c_2(\mu)O_2^u(\mu))   \nonumber \\
&+& V_{cb}V_{cq}^*(c_1(\mu)O_1^c(\mu)+c_2(\mu)
O_2^c(\mu))+\cdots \Big]+{\rm h.c.},
\eeq
where $q=d,s$, and
\beq \label{O12}
&& O_1^u= \bar c\gamma_\mu(1-\gamma_5)b\,\,\bar q\gamma^\mu(1-\gamma_5)u, 
\qquad\quad
O_2^u = 
\bar q\gamma_\mu(1-\gamma_5)b\,\,\bar c\gamma^\mu(1-\gamma_5)u \,.
\eeq
The scale and scheme dependence of the Wilson coefficients $c_{1,2}(\mu)$ are 
canceled out by the corresponding dependence in the matrix element of the
four-quark operators $O_{1,2}$. That is, the four-quark operators 
in the effective theory have to
be renormalized at the same scale $\mu$ and evaluated using the same
renormalization scheme as that for the Wilson coefficients. Schematically,
we can write $\langle {\cal H}_{\rm eff}\rangle=c(\mu)\langle 
O(\mu)\rangle=c(\mu)
g(\mu)\langle O\rangle_{\rm tree}=c^{\rm eff}\langle O\rangle_{\rm tree}$,
where the effective Wilson coefficients $c_i^{\rm eff}$ are renormalization
scale and scheme independent. Then the factorization approximation or the
quark model is applied to evaluate the hadronic matrix elements of the 
operator $O$ at tree level.
The explicit expression for $g(\mu)$,
the perturbative corrections to the four-quark operators renormalized 
at the scale $\mu$, has been 
calculated in the literature \cite{Buras,Ali}. To the next-to-leading order 
(NLO) precision, we have \cite{CT}
\footnote{The effective Wilson coefficients given in (\ref{c12}) are 
derived from $c_i(\mu)$ at $\mu=m_b$ to the NLO \cite{Cheng}. Nevertheless,
it is not difficult to explicitly check the scale and scheme independence
of $c_i^{\rm eff}$. For example, the authors of \cite{Ali}
obtain $c_1^{\rm eff}=1.160$ and $c_2^{\rm eff}=-0.334$ at $\mu=2.5$ GeV.
Therefore, $c_i^{\rm eff}$ are very insensitive to the chosen $\mu$ scale,
as it should be. It is known that the Wilson coefficient $c_2$ at the NLO:
$c_2=-0.185$ in the naive dimension regularization scheme and $c_2=-0.228$ 
in the 't Hooft-Veltman scheme \cite{Buras}, deviates substantially from the 
leading-order value $c_2=-0.308$ at $\mu=m_b(m_b)$. However, 
the resultant $c_2^{\rm eff}$ is scheme independent and its value is 
close to the leading-order one.}
\beq \label{c12}
c_1^{\rm eff}=1.149\,, \qquad \quad c_2^{\rm eff}=-0.325\,.
\eeq

Replacing $c_i$ by $c_i^{\rm eff}$ and using $m_b=(4.85\pm 0.25)$ GeV, 
$(m_c/m_b)^2=$0.089, 
$|V_{cb}|=0.039$, $G_{B_u,B_d}=0.366\,{\rm GeV}^2$, 
$G_{B_s}=0.381\,{\rm GeV}^2$ \cite{PDG},
$G_{\Lambda_b}=0$, $K_{B_u,B_d}\approx K_{\Lambda_b}\approx 
0.4\,{\rm GeV}^2$, $K_{B_s}=1.02\times K_{B_u,B_d}$ \cite{HL}
together with the nonleptonic
inclusive results to the next-to-leading order \cite{Bagan},
we find numerically
\beq   \label{num}
&& \Gamma_{\rm SL}(B\to e\bar\nu X) = (4.18^{+1.20}_{-0.99})
\times 10^{-14}\,{\rm GeV}, 
\nonumber \\
&& \Gamma_{\rm SL}(\Lambda_b\to e\bar\nu X)= (4.32^{+1.24}_{-1.01})
\times 10^{-14}\,{\rm GeV},
\nonumber \\
&& \Gamma(B) = \Gamma_{\rm NL}(B)+2.24\,\Gamma_{\rm SL}(B\to e\bar\nu X)=
(3.61^{+1.04}_{-0.84})\times 10^{-13}\,{\rm GeV},   \nonumber \\
&& \Gamma(\Lambda_b) = \Gamma_{\rm NL}(\Lambda_b)+2.24\,\Gamma_{\rm SL}
(\Lambda_b\to e\bar\nu X)=(3.65^{+1.04}_{-0.85})\times 10^{-13}\,{\rm GeV}.   
\eeq
It follows that the lifetime ratios of the $H_b$ hadrons are 
\begin{eqnarray}\label{taucrude}
   \frac{\tau(B^-)}{\tau(B_d)} &=& 1 + O(1/m_b^3) \,, \nonumber\\
   \frac{\tau(B_s)}{\tau(B_d)} &=& 1.0005 + O(1/m_b^3) \,, \nonumber\\
   \frac{\tau(\Lambda_b)}{\tau(B_d)} &=& 
   0.99 + O(1/m_b^3) \,.
\end{eqnarray}
Note that $\tau(B_s)$ here refers to the average lifetime of the two CP
eigenstates of the $B_s$ meson. It is evident that the $1/m_b^2$
corrections are too small to explain the shorter lifetime of the
$\Lambda_b$ relative to that of the $B_d$.
To the order of $1/m_b^3$, the nonspectator effects due to Pauli 
interference and
$W$-exchange parametrized in terms of the hadronic parameters~\cite{NS}:
$B_1$, $B_2$, $\varepsilon_1$, $\varepsilon_2$,
$\tilde B$, and $r$ (see below), may contribute significantly to
lifetime ratios due to a phase-space enhancement by a factor of $16\pi^2$.
The four-quark operators relevant to inclusive nonleptonic $B$ decays are
\begin{eqnarray}\label{4qops}
   O_{V-A}^q &=& \bar b_L\gamma_\mu q_L\,\bar q_L\gamma^\mu b_L
    \,, \nonumber\\
   O_{S-P}^q &=& \bar b_R\,q_L\,\bar q_L\,b_R \,, \nonumber\\
   T_{V-A}^q &=& \bar b_L\gamma_\mu t^a q_L\,
\bar q_L\gamma^\mu  t^a b_L \,, \nonumber\\
   T_{S-P}^q &=& \bar b_R\,t^a q_L\,\bar q_L\, t^ab_R \,,
\end{eqnarray}
where
$q_{R,L}={1\pm\gamma_5\over 2}q$ and $t^a=\lambda^a/2$ with $\lambda^a$
being the Gell-Mann matrices. For the matrix elements of these 
four-quark operators between $B$ hadron states,
we follow \cite{NS} to adopt the following definitions:
\beq \label{parameters}
{1\over 2m_{ B_q}}\la \bar B_q|\Ova|\bar B_q\ra &&\equiv {f^2_{B_q} m_{B_q}
\over 8}B_1\,, \nonumber\\
{1\over 2m_{B_q}}\la \bar B_q|\Osp|\bar B_q\ra &&\equiv {f^2_{B_q} 
m_{B_q}\over 8}B_2\,,\nonumber\\
{1\over 2m_{B_q}}\la \bar B_q|\Tva|\bar B_q\ra &&\equiv {f^2_{B_q} 
m_{B_q}\over 8}\varepsilon_1\,, \nonumber\\
{1\over 2m_{B_q}}\la \bar B_q|\Tsp|\bar B_q\ra &&\equiv {f^2_{B_q} 
m_{B_q}\over 8}\varepsilon_2\,,\nonumber\\
{1\over 2m_{\Lambda_b}}\la \Lambda_b |\Ova|\Lambda_b \ra 
&&\equiv -{f^2_{B_q} m_{B_q}\over 48}r\,,\nonumber\\
{1\over 2m_{\Lambda_b}}\la \Lambda_b |\Tva|\Lambda_b \ra
&&\equiv -{1\over 2} (\tilde B+{1\over 3})
{1\over 2m_{\Lambda_b}}\la \Lambda_b |\Ova|\Lambda_b \ra \,,
\eeq
where $f_{B_q}$ is the $B_q$ meson decay constant defined by
\begin{equation}
\langle 0|\bar q\gamma_\mu \gamma_5 b|\bar B_q(p)\rangle =if_{B_q} p_\mu \,.
\end{equation}
Under the factorization approximation, $B_i=1$ and $\varepsilon_i=0$, and
under the valence quark approximation $\tilde B=1$ \cite{NS}.

To the order of $1/m_b^3$, we find that the $B$-hadron lifetime ratios are 
given by
\begin{eqnarray}\label{ratios}
\frac{\tau(B^-)}{\tau(B^0_d)} &=& 1 +
   \Bigl( {f_B\over 185~{\rm MeV}} \Big)^2 \Big( 0.043 B_1 + 0.0006 B_2
    - 0.61 \varepsilon_1 + 0.17 \varepsilon_2 \Big) \,, \nonumber \\
\frac{\tau (B^0_s)}{\tau(B^0_d)}
&=& 1+ \Big(\frac{f_B}{185~{\rm MeV}}\Bigr)^2
(-1.7\times 10^{-5}\,B_1+1.9\times 10^{-5}\, B_2 \,-0.0044 
\varepsilon_1\, +0.0050\, \varepsilon_2)  \,, \nonumber\\
   \frac{\tau(\Lambda_b)}{\tau(B^0_d)} &=& 0.99
   +  \Bigl( {f_B\over 185~{\rm MeV}} \Big)^2 \Big[ -0.0006 B_1
    + 0.0006 B_2   \nonumber \\
&&   - 0.15 \varepsilon_1 + 0.17 \varepsilon_2
    - (0.014 + 0.019 \widetilde B) r \Big] \,.
\end{eqnarray}
The above results are similar to that given in \cite{NS}.
We see that the coefficients of the color singlet--singlet
operators are one to two orders of magnitude smaller than those of
the color octet--octet operators. This implies that
even a small deviation from the factorization approximation 
$\varepsilon_i=0$ can have a sizable impact on the lifetime ratios.
It was argued in \cite{NS} that the unknown nonfactorizable contributions
render it impossible to make reliable estimates on the magnitude of the
lifetime ratios and even the sign of corrections. That is, the theoretical 
prediction for $\tau(B^-)/\tau(B_d)$ is not necessarily larger than unity.
In the next section we will apply the QCD sum rule method to estimate 
the aforementioned hadronic parameters, especially $\varepsilon_i$.

\section{The QCD sum rule calculation}
    In this section we will employ the method of QCD sum rules within the
framework of HQET. Since the $b$ quark is treated as a static quark with 
an infinite quark mass in HQET and since HQET is a low-energy effective
theory, it is natural to regard the matrix elements of (\ref{parameters})
as defined at the scale $m_b$ and to evaluate the corresponding hadronic
matrix elements in HQET at a scale $\mu<m_b$. Indeed,
it has been argued that the estimate of hadronic matrix elements of
four-quark operators using the factorization hypothesis for mesons and
the quark model for baryons becomes more reliable if the operators are 
renormalized at a typical hadronic scale \cite{BS94}. 
In the sum rule approach, the correlation
function (or the so-called Green function), 
within the QCD framework, can be expanded as a series of 
operators $O_n(\mu)$ multiplied by the Wilson coefficients 
$C_n(-2\omega/\mu,g_s)/\omega^n$,
where $\omega$ is an external momentum 
flowing in (or out) the correlation function and $\mu$ is the factorization
scale that separates the long-distance part $O_n$ from the short-distance one
$C_n$. The quality of the convergence
of the OPE series is controlled by the value of 
the external momentum $\omega$.
The factorization scale $\mu$ cannot be chosen too small,
otherwise the strong coupling constant $\alpha_s$ would be so large that
Wilson coefficients cannot be perturbatively calculated.
Four-quark operators are sometimes renormalized at a typical scale 
$\mu_h \approx 0.67~{\rm GeV}$, corresponding to the coupling
constant $\alpha_s^{\over {\rm MS}}(\mu_h)\sim {\cal O}(1)$. 
However, such a scale is not quite 
suitable for the sum rule calculation. Instead we choose $\mu=1~{\rm GeV}$
as the lowest possible factorization scale in the ensuing study. 
After summing over the logarithmic dependence
$\alpha_s^m {\rm ln}^m (-2\omega/\mu)$ by the renormalization-group method, 
one obtains the nonperturbative quantity $X(\mu)$ which can be extracted from
the correlation function in the following form
\begin{eqnarray}
X(\mu)\sim \sum_n { C_n(1, g_s(-2\omega))\over \omega^n} 
\Bigl( {\alpha_s(\mu)\over \alpha_s(-2\omega)} \Bigr)^{\gamma_n-\sum_j 
\gamma_j}O_n(\mu)\,, 
\end{eqnarray}
where $\gamma_n$ are the anomalous dimensions of $O_n$ and $\gamma_j$ 
the anomalous dimensions of the currents appearing in the correlation 
function. As noted above, the four-quark operators in Eq.~(\ref{parameters}) 
are defined at the scale $\mu=m_b$. In HQET where the $b$ quark is
treated as a static quark,
we can use the renormalization group equation to express 
them in terms of the operators renormalized at a scale
 $ \Lambda_{\rm QCD}\ll \mu\ll m_b$. These operators have
the hybrid anomalous dimensions \cite{SV1,SV2,PW} and
their renormalization-group
evolution is determined by the anomalous dimensions in HQET.
The operators $O_{V-A}^q$ and $T_{V-A}^q$,
and similarly $O_{S-P}^q$ and $T_{S-P}^q$, mix under renormalization.
In the leading logarithmic approximation, the renormalization-group equation
of the operator pair $(O,T)$ governed
by the hybrid anomalous dimension matrix reads 
\footnote{One of the off-diagonal anomalous dimension matrix elements 
in Eq.~(\ref{rge}) has a sign opposite to that obtained in \cite{NS},
but the final result in Eq.~(\ref{diagp}) is in full agreement with the
results derived there.}
\begin{equation}\label{rge}
\frac{d}{dt}
\left(
   \begin{array} {cc} \phantom{ \bigg[ } O \\
   T\end{array} \right)
    = {3\alpha_s\over 2\pi}\,\left(
   \begin{array} {cc} \phantom{ \bigg[ } C_F & -1 \\
    \displaystyle -{C_F\over 2 N_c}~ & \displaystyle ~{1\over 2 N_c}
   \end{array} \right)
\left(
   \begin{array} {cc} \phantom{ \bigg[ } O \\
   T\end{array} \right)
 \,,
\end{equation}
where $t = \frac{1}{2} \ln(Q^2/\mu^2)$, 
$C_F = (N_c^2-1)/2 N_c$, and effects of penguin operators induced
from evolution have been neglected.

The solution to the evolution equation
Eq.~(\ref{rge}) has the form
\begin{equation}\label{diag}
\left(
   \begin{array} {cc} O \\
   T\end{array} \right)_{Q}
    = \left(
   \begin{array} {cc} \phantom{ \bigg[ } \frac{8}{9}~ & \frac{2}{3} \\
    -\frac{4}{27}~ &
\frac{8}{9}
   \end{array} \right)
\left(
   \begin{array} {cc} \phantom{ \bigg[ } L_{Q}^{9/(2\beta_0)}~ & 0 \\
    0~ & 1
   \end{array} \right)
 {\bf D_\mu}\,,
\end{equation}
where
\begin{equation}\label{d}
{\bf D_\mu}=
\left(
   \begin{array} {cc} D_1 \\
   D_2\end{array} \right)_\mu
    = \left(
   \begin{array} {cc} \phantom{ \bigg[ } O-\frac{3}{4}T \\
   \frac{1}{6}O+T\end{array} \right)_\mu
 \,,
\end{equation}

\begin{equation}
   L_Q = {\alpha_s(\mu)\over\alpha_s(Q)} \,,
\end{equation}
and $\beta_0=\frac{11}{3}\,N_c-\frac{2}{3}\,n_f$ is the leading-order 
expression of the $\beta$-function with $n_f$ being the number of light
quark flavors. The subscript $\mu$ in Eq.~(\ref{d}) and in what follows 
denotes the renormalization point of the operators.
Given the evolution equation (\ref{diag}) for the four-quark
operators, we see that the hadronic parameters 
$B_i$ and $\varepsilon_i$ normalized at the scale $m_b$ are related to that
at $\mu=1$~GeV by
\begin{equation}\label{diagp}
\left(
   \begin{array} {cc} B_i \\
   \varepsilon_i\end{array} \right)_{m_b}
    = \left(
   \begin{array} {cc} \phantom{ \bigg[ } \frac{8}{9}~ & \frac{2}{3} \\
    -\frac{4}{27}~ &
\frac{8}{9}
   \end{array} \right)
\left(
   \begin{array} {cc} \phantom{ \bigg[ } L_{m_b}^{9/(2\beta_0)}~ & 0 \\
    0~ & 1
   \end{array} \right)
     \left(
   \begin{array} {cc} \phantom{ \bigg[ } B_i-\frac{3}{4}\varepsilon_i \\
   \frac{1}{6}B_i+ \varepsilon_i\end{array} \right)_{\rm \mu=1~GeV}
 \,,
\end{equation}
and hence
\begin{eqnarray}\label{lowpara}
   B_i(m_b) &\simeq& 1.54 B_i(\mu)
    - 0.41\varepsilon_i(\mu) \,, \nonumber\\
   \varepsilon_i(m_b) &\simeq& - 0.090 B_i(\mu)+1.07\varepsilon_i(\mu) \,,
\end{eqnarray}
with $\mu=1$ GeV, where uses have been made of
$\alpha_s(m_{\rm Z})=0.118$, 
$\Lambda^{(4)}_{\over{\rm MS}}=333~{\rm MeV}$, $m_b=4.85~{\rm GeV}$,
$m_c=1.45$ GeV and 
$$\alpha_s(Q)={4\pi \over \beta_0 {\rm ln}{Q^2\over\Lambda^2} }
\left( 1-{2\beta_1\over \beta^2_0}\,
{{\rm ln} ({\rm ln} {Q^2\over \Lambda^2})\over  
{\rm ln}{Q^2\over \Lambda^2} }\right) 
$$
to the next-to-leading order with $\beta_1=51-{19\over 3}n_f$.
The above results (\ref{lowpara}) indicate that renormalization effects 
are quite significant. 

It is easily seen from Eqs.~(\ref{diag})
and (\ref{d})
that the normalized operator $D_1$ (or $D_2$) is simply multiplied by 
$L_{Q}^{9/(2\beta_0)}$ (or 1) when it evolves from a renormalization point 
$\mu$
to another point $Q$. In what follows, we will apply this  property to derive
the renormalization-group improved QCD sum rules for $D_j$ at the typical 
scale $\mu=1$~GeV.
We define the new four-quark matrix elements as follows
\beq
{1\over 2m_{ B_q}}\la \bar B_q|D_j^{(i)}(\mu)|\bar B_q\ra
\equiv {f^2_{B_q} m_{B_q}\over 8}\, d_j^{(i)}(\mu),
\eeq
where the superscript $(i)$ denotes $(V-A)$ four-quark operators for $i=1$
and $(S-P)$ operators for $i=2$, and $d_j^{(i)}$ satisfy
\begin{equation}
\left(
   \begin{array} {cc} d_1^{(i)} \\
   d_2^{(i)}\end{array} \right)_\mu
    = \left(
   \begin{array} {cc} \phantom{ \bigg[ } B_i-\frac{3}{4}\varepsilon_i \\
   \frac{1}{6}B_i+\varepsilon_i\end{array} \right)_\mu
 \,.
\end{equation}

Since the terms linear in
four-quark matrix elements are already of order $1/m_b^3$, we  only need the
relation between the full QCD field $b(x)$ and the HQET field $h^{(b)}_v(x)$ 
to the zeroth order in $1/m_b$:
$b(x) = e ^{-im_b v\cdot x}\, \{h^{(b)}_v(x) + {\cal O}(1/m_b)\}$.
Therefore, in analogue to Eq.~(\ref{4qops}),
we define the relevant four-quark operators in HQET as
\begin{eqnarray}
 O_{V-A}^v &=& \bar h^{(b)}_{vL}\gamma_\mu q_L\,
\bar q_L\gamma^\mu h^{(b)}_{vL} \,, \nonumber\\
   O_{S-P}^v &=& \bar h^{(b)}_{vR}\,q_L\,\bar q_L\,h^{(b)}_{vR} \,,\nonumber\\
   T_{V-A}^v &=& \bar h^{(b)}_{vL}\gamma_\mu t^a q_L\,
\bar q_L\gamma^\mu  t^a h^{(b)}_{vL} \,, \nonumber\\
   T_{S-P}^v &=& \bar h^{(b)}_{vR}\,t^a q_L\,\bar q_L\,t^a h^{(b)}_{vR} \,.
\end{eqnarray}
The corresponding hadronic matrix elements of these four-quark operators 
are parametrized by
\beq\label{4qopshqet}
{1\over 2}\la \bar B(v)|O_{V-A}^v|\bar B(v)\ra
&\equiv& {F^2(m_b)\over 8} B_1^v(\mu), \nonumber\\
{1\over 2}\la \bar B(v)|O_{S-P}^v|\bar B(v)\ra
&\equiv& {F^2(m_b) \over 8} B_2^v(\mu),  \nonumber\\
{1\over 2}\la \bar B(v)|T_{V-A}^v|\bar B(v)\ra
&\equiv& {F^2(m_b) \over 8} \varepsilon_1^v(\mu), \nonumber\\
{1\over 2}\la \bar B(v)|T_{S-P}^v|\bar B(v)\ra
&\equiv& {F^2(m_b) \over 8} \varepsilon_2^v(\mu),
\eeq
where the heavy-flavor-independent decay constant $F$ defined in the heavy 
quark limit is given by
\beq
\langle 0|\bar q\gamma^\mu\gamma_5 h^{(b)}_v|\bar B(v)\rangle
=iF(\mu) v^\mu\,.
\eeq
The decay constant $F(\mu)$ depends on the scale $\mu$ at which the effective
current operator is renormalized and it is related to the scale-independent 
decay constant $f_B$ of the $B$ meson by
\beq
F(m_b)=f_B\,\sqrt{m_B}.
\eeq
Notice that $F$ in Eq.~(\ref{4qopshqet})
is chosen to be normalized at the scale $m_b$.

To complete the aim of obtaining the matrix elements of four-quark
operators, we apply the method of QCD sum rules~\cite{SVZ}.
We consider the following three-point correlation function
\beq \label{corr}
\Pi^{D_j^{v(i)}}_{\alpha,\beta}(\omega,\omega')=i^2\int dx\, dy\,
e^{i\omega v\cdot x-i\omega' v\cdot y}
\la 0|T\{[\bar q(x)\Gamma_\alpha h^{(b)}_v(x)]\, D_j^{v(i)}(0)\,
[\bar q(y)\Gamma_\beta h^{(b)}_v(y)]^\dagger\}|0\ra
\,,
\eeq
of the operator $D_j^{(i)}$ defined in Eq.~(\ref{d}), where
$\Gamma_\alpha=\gamma_\alpha\gamma_5$.
However, this current interpolates not only the heavy mesons with
quantum number $J^P=0^-$ but also that with quantum number $J^P=1^+$.
Therefore, we need to decompose $\Gamma_{\alpha}$ into
$\Gamma_{\alpha}=\Gamma_{\alpha}^{AV}
-v_{\alpha}\Gamma^{PS}$, with 
$\Gamma_{\alpha}^{AV}=(\gamma+v)_{\alpha}\gamma_5$ for $J^P=1^+$
and $\Gamma^{PS}=\gamma_5$ for $J^P=0^-$.
As a consequence, $\Pi^{D_j^{v(i)}}_{\alpha\beta}$ is recast to 
\beq \label{corrp}
\Pi^{D_j^{v(i)}}_{\alpha\beta} =(-g_{\alpha\beta}+v_\alpha v_\beta)\Pi^{AV}_{
D_j^{v(i)}}+v_\alpha v_\beta \Pi^{PS}_{D_j^{v(i)}},
\eeq
where
\beq \label{corrpp}
(-g_{\alpha\beta}+v_\alpha v_\beta)\Pi^{AV}_{D_j^{v(i)}} &=&
i^2\int dx\, dy\, e^{i\omega v\cdot x-i\omega' v\cdot y}  \nonumber \\
&\times& \la 0|T\{[\bar q(x)\Gamma^{AV}_\alpha h^{(b)}_v(x)]\, D_j^{v(i)}(0)\,
[\bar q(y)\Gamma_\beta^{AV} h^{(b)}_v(y)]^\dagger\}|0\ra \,, \nonumber\\
\Pi^{PS}_{D_j^{v(i)}} &=& i^2\int dx\, dy\,e^{i\omega v\cdot x-i\omega' 
v\cdot y}   \nonumber \\
&\times& \la 0|T\{[\bar q(x)\Gamma^{PS} h^{(b)}_v(x)]\, D_j^{v(i)}(0)\,
[\bar q(y)\Gamma^{PS} h^{(b)}_v(y)]^\dagger\}|0\ra \,.
\eeq
In deriving Eq.~(\ref{corrp}) we have applied the relations
\beq
i^2\int dx\, dy\,
e^{i\omega v\cdot x-i\omega' v\cdot y}
\la 0|T\{[\bar q(x)\Gamma^{AV}_\alpha h^{(b)}_v(x)]\, D_j^{v(i)}(0)\,
[\bar q(y)\Gamma^{PS} h^{(b)}_v(y)]^\dagger\}|0\ra
=0 
\eeq 
and
\beq
i^2\int dx\, dy\,
e^{i\omega v\cdot x-i\omega' v\cdot y}
\la 0|T\{[\bar q(x)\Gamma^{PS} h^{(b)}_v(x)]\, D_j^{v(i)}(0)\,
[\bar q(y)\Gamma_\beta^{AV} h^{(b)}_v(y)]^\dagger\}|0\ra 
=0 \,.
\eeq
Note that only the correlation function $\Pi^{PS}$ is relevant to our purpose.
It can be written in the double dispersion
relation form
\beq
\Pi^{PS}_{D_j^{v(i)}}(\omega,\omega')=\int\int {ds\over s-\omega}\,
{ds'\over s'-\omega'}\, \rho^{D_j^{v(i)}}
\,.
\eeq

The results of the QCD sum rules can be obtained in the following way.
On the phenomenological side, which is the sum of the relevant
hadron states, this correlation function can be written as
\beq
\Pi^{PS}_{D_j^{v(i)}}(\omega,\omega')=
\frac{F^2(m_b)F^2(\mu)d_j^{(i)}}
{16(\bar \Lambda -\omega)(\bar \Lambda -\omega')}+\cdots \,,
\eeq
where $\bar\Lambda$ is the binding energy of the heavy meson in the
heavy quark limit and ellipses denote resonance contributions.
On the theoretical side, the correlation function $\Pi^{PS}$ can be 
alternatively calculated in terms of quarks and gluons using the standard
OPE technique.  Then we equate the results
on the phenomenological side with that on the theoretical side. However, since
we are only interested in the properties of the ground state at hand,
e.g., the $B$ meson,
we shall assume that contributions from excited states
(on the phenomenological side)
are approximated by the spectral density on the theoretical side of the
sum rule, which starts from some thresholds (say, $\omega_{i,j}$ in this 
study).
To further improve the final result under consideration,
we apply the Borel transform to both external variables $\omega$ and
$\omega'$. After the Borel transform~\cite{SVZ}, 
\beq
{\bf B}[\Pi^{PS}_{D_j^{v(i)}}(\omega,\omega')]=
\lim_{{\scriptstyle m\to \infty \atop\scriptstyle -\omega'\to \infty}
\atop\scriptstyle {-\omega'\over mt'}\ {\rm fixed}}
\lim_{{\scriptstyle n\to \infty \atop\scriptstyle -\omega\to \infty}
\atop\scriptstyle {-\omega\over nt}\ {\rm fixed}}
{1\over n!m!}(-\omega')^{m+1} [{d\over d\omega'}]^m
(-\omega)^{n+1} [{d\over d\omega}]^n \Pi^{PS}_{D_j^{v(i)}}(\omega,\omega')\,,
\eeq
the sum rule gives
\beq
&&\frac{F^2({m_b}) F^2(\mu)}{16} e^{- \bar\Lambda/t_1}
e^{-\bar\Lambda/t_2} d_j^{(i)}
=\int_0^{\omega_{i,j}} ds\int_0^{\omega_{i,j}} ds'
e^{-(s/t_1 + s'/t_2)}\rho^{\rm QCD}\,,
\eeq
where $\omega_{i,j}$ is the threshold of the excited states and $\rho^{\rm 
QCD}$ is the spectral density on the theoretical side of the sum rule.
Because the Borel windows are symmetric in variables $t_1$ and $t_2$,
it is natural to choose $t_1=t_2$. However, unlike the case of the
normalization of the Isgur-Wise function at zero recoil, where the Borel
mass is approximately twice as large as that in the corresponding two-point
sum rule~\cite{Neubert2},
in the present case of the three-point sum rule at hand, we find
that the working Borel windows can be chosen as the same as that 
in the two-point sum rule since in our analysis the output results
depend weakly on the Borel mass. Therefore, we choose $t_1=t_2=t$. 
 By the renormalization group technique, the logarithmic dependence
$\alpha_s\ln (2t/\mu)$ can be summed over to produce a factor like
$[\alpha_s(\mu)/\alpha_s(2t)]^\gamma$. 
After some manipulation we obtain the sum rule results:
\begin{eqnarray}\label{rule1}
&&\frac{F^2({m_b}) F^2(\mu)}{16} e^{-2 \bar\Lambda/t}
\left(
   \begin{array} {cc} \phantom{ \bigg[ } d_1^{v(i)} \\
   d_2^{v(i)}\end{array} \right)_\mu  \nonumber\\
=&&
\Biggl( {\alpha_s(2t)\over \alpha_s(\mu)} \Biggr)^{4\over\beta_0}
\Biggl( {1-2\delta{\alpha_s(2t)\over\pi}\over 1-2\delta{\alpha_s(\mu)
\over\pi}}\Biggr)
 \left(
   \begin{array} {cc} L_{t}^{-9/(2\beta_0)}~ & 0 \\
    0~ & 1
   \end{array} \right)
 \left(
   \begin{array} {cc} \phantom{ \bigg[ } {\rm OPE}_{B_{i,1}}
   -\frac{3}{4}\,{\rm OPE}_{\varepsilon_{i,1}} \\
    \frac{1}{6}\, {\rm OPE}_{B_{i,2}}+{\rm OPE}_{\varepsilon_{i,2}}
   \end{array}
   \right)_t \,,
\end{eqnarray}
where
\begin{eqnarray}\label{rule2}
&&{\rm OPE}_{B_{i,j}}\simeq \frac{1}{4}({\rm OPE})^2_{2pt;i,j} \,, \nonumber\\
&&
{\rm OPE}_{\varepsilon_{1,j}} \simeq -\frac{1}{16}\Biggl[-
\frac{ \langle \bar q g_s\sigma\cdot G q\rangle}{8\pi^2} t (1-e^{-\omega_{1,j} 
/t})+\frac {\langle \alpha_s G^2\rangle} {16\pi^3} t^2 (1-e^{-\omega_{1,j} 
/t})^2 \Biggr] \,, \nonumber\\
&& {\rm OPE}_{\varepsilon_{2,j}} \simeq {\cal O}(\alpha_s)\,,
\end{eqnarray}
with
\begin{eqnarray} \label{2pt}
({\rm OPE})_{2pt;i,j}
=&&
\frac{1}{2}  \biggl\{
\int_0^{\omega_{i,j}} ds\ s^2 e^{-s/t}{3\over\pi^2}
\biggl[1+{\alpha_s\over\pi}
\Bigl({17\over 3}+{4\pi^2\over 9}-2\ln {s\over t}\Bigr)
\biggr]\nonumber\\
&&-\Bigl(1+{2\alpha_s\over\pi}\Bigr)\langle\bar qq\rangle
+{\langle\bar qg_s \sigma\cdot Gq\rangle\over 16 t^2}
\biggr\} \,.
\end{eqnarray}
For reason of consistency,
in the following numerical analysis we will neglect the finite 
part of radiative one loop corrections in OPE$_{B_{i,j}}$ and 
OPE$_{\varepsilon_{i,j}}$ (and in Eq. (\ref{F})).
The parameter $\delta$ in (\ref{rule1}) is some combination of the $\beta$
functions and anomalous dimensions (see Eq.~(4.2) of \cite{BB}) and is 
numerically equal to $-0.23$. The relevant parameters normalized at the 
scale $t$ are related to those at $\mu$ by~\cite{bbbd,Neubert2,yang1}
\beq\label{RGevo}
&&F(2t)=F(\mu)\Bigl( {\alpha_s(2t)\over \alpha_s(\mu)} \Bigr)^{-2/\beta_0}
{1-\delta{\alpha_s(\mu)\over\pi} \over 1-\delta{\alpha_s(2t)\over\pi}}\,,
\nonumber\\
&&\langle \bar qq\rangle_{2t} =\langle \bar qq\rangle_\mu \cdot
\Bigl( {\alpha_s(2t)\over \alpha_s (\mu)} \Bigr)^{-4/\beta_0}\,,\nonumber\\
&&\langle g_s\bar q\sigma\cdot Gq\rangle_{2t}=\langle g_s\bar q\sigma\cdot 
Gq\rangle_\mu
\cdot  \Bigl( {\alpha_s(2t)\over \alpha_s(\mu)} \Bigr)^{2/(3\beta_0)}
\,,\nonumber\\
&&\langle \alpha_s G^2 \rangle_{2t}= \langle \alpha_s G^2 \rangle_\mu\,,
\eeq
where 
$\langle\cdots \rangle$ stands for $\langle 0| \cdots |0\rangle$. 
In the calculation of the correlation function, we have also used
the fixed-point gauge (the Fock-Schwinger gauge) $x^\mu A_\mu(x)=0$ 
with $A_\mu$ being an external gluon field. Under this gauge, the 
generalized quark propagator in the external gluon field reads
\begin{eqnarray}
S^{ab}_{q(ij)}(0,x)=&&\int {d^4p\over(2\pi)^4}e^{ip\cdot x}
\left[
{i\delta^{ab}\over \not p-m_q} \right.
+ {i\over 4}{\lambda^n_{ab}\over 2}g_s G_{\mu\nu}^n(0)
{\sigma^{\mu\nu}(\not p + m_q) + (\not p + m_q)
\sigma^{\mu\nu}\over (p^2-m_q^2)^2}\nonumber\\
&&\left. -{iG_{\mu\nu}^n(0)\lambda^n_{ab}\over 4}g_s x^\nu
({1\over \not p-m_q}\gamma^\mu {1\over \not p-m_q})
\right]_{ij}\nonumber\\
&&+:q^a_i(0) \bar q^b_j(0):+ x_\mu:q^a_i(0)(D^\mu\bar q^b_j(0)):+
{x_\mu x_\nu\over 2!}:q^a_i(0)(D^\mu D^\nu \bar q^b_j(0)):\nonumber\\
&&+\dots \,,
\end{eqnarray}
where $a$ and $b$ are the color indices, $i$ and $j$ the Lorentz
indices.

Let us explain the results obtained in Eqs.~(\ref{rule1}) and (\ref{rule2}).
OPE$_{B_i}$ is obtained by substituting $D_j^{v(i)}$ by
$O^v$ in $\Pi^{PS}_{D_j^{v(i)}}$ [cf. Eq.~(\ref{corrpp})] and it can be 
approximately factorized
as the product of (OPE)$_{2pt;i,j}$ with itself, which is the same
as the theoretical part in the two-point $F(\mu)$ sum
rule~\cite{Neubert2,BB,bbbd}.  In the series of (OPE)$_{2pt;i,j}$, we have
neglected the contribution proportional to $\langle \bar qq\rangle^2$.
(More precisely, it is equal to $\alpha_s \langle \bar qq\rangle^2 \pi/324$;
see Ref.~\cite{Neubert2}.) Nevertheless,
the result of (OPE)$_{B_i}$ in Eq.~(\ref{rule2}) 
is reliable up to dimension six, as the contributions from the $\langle \bar 
qq\rangle^2$ terms in (OPE)$_{2pt;i,j}$ are much smaller than the term
$(1+\alpha_s/\pi)^2 \langle \bar qq \rangle^2/16$
that we have kept [see Eq.~(\ref{2pt})]. Note that in (OPE)$_{B_i}$ the 
contribution involving the gluon condensate is proportional to the light
quark mass and hence can be neglected. 
Likewise, OPE$_{\varepsilon_i}$ is the theoretical side of the sum rule, and
it is obtained by substituting
$D_j^{v(i)}$ by $T^v$ in Eq.~(\ref{corrpp}). To the order of dimension-five,
the main contributions to OPE$_{\varepsilon_i}$ are depicted in Fig.~1.
Here we have neglected the dimension-6 four-quark condensate
of the type $\langle \bar q\Gamma\lambda^a q\ \bar q\Gamma\lambda^a 
q \rangle$. It's contribution is much less than that from 
dimension-five or dimension-four condensates and
hence unimportant (see~\cite{Chern} for similar discussions). It should be
emphasized that nonfactorizable contributions to the parameters $B_i$ arise
mainly from the $O^v-T^v$ operator mixing.

At this
point, it is useful to compare our analysis with the similar QCD sum rule
studies in \cite{Chern} and \cite{BLLS}. First,
Chernyak \cite{Chern} used the chiral interpolating
current for the $B$ meson, so that all light quark fields in his correlators
are purely left-handed. As a result, there are no quark-gluon mixed 
condensates as
these require the presence of both left- and right-handed light quark fields.
The gluon condensate contribution enters into the $\varepsilon_1$ sum rule
with an additional factor of 4 in comparison with ours.
Second, our results for OPE$_{\varepsilon_i}$ are very different from that 
obtained by
Baek {\it et al}.~\cite{BLLS}. The reason is that they calculated 
the full $\Pi^{\varepsilon_i,\alpha}_\alpha$ (obtained by replacing 
$D_j^{v(i)}$ by 
$T^v$ in Eq.~(\ref{corr})) rather than  the pseudoscalar part
of $\Pi^{\varepsilon_i,\alpha}_\alpha$. 
Therefore, their results are mixed with the $1^+$ to $1^+$
transitions. Also a subtraction of the contribution from excited states 
is not carried out in \cite{BLLS} for the three-point correlation function, 
though it is justified to do so for two-point correlation functions.
Indeed, in the following analysis, one will find that after
subtracting the contribution from excited states, the contributions
of OPE$_{\varepsilon_i}$ are largely suppressed. Furthermore,
as in the study of the $B$ meson decay constant~\cite{Neubert2}, we find that
the renormalization-group effects are very important in the sum rule analysis. 
Consequently, there is no much difference between the resulting values
of $\varepsilon_1$ and $\varepsilon_2$. Moreover, $\varepsilon_i$ at 
$\mu=m_b$ are largely enhanced by renormalization-group effects.

The value of $F$ in Eq.~(\ref{rule1}) can be substituted by
\begin{eqnarray}\label{F}
F^2(\mu)e^{-\bar\Lambda/t}=&&
\biggl[{\alpha_s(2t)\over \alpha_s(\mu)}\biggr]^{4\over\beta}
\biggl[{1-2\delta{\alpha_s(2t)\over\pi}\over 1-2\delta{\alpha_s(\mu)\over\pi}}
\biggr]  \biggl\{
\int_0^{\omega_0} ds\ s^2 e^{-s/ t}{3\over\pi^2}
\biggl[1+{\alpha_s(2t)\over\pi}
\Bigl({17\over 3}+{4\pi^2\over 9}-2\ln {s\over t}\Bigr)
\biggr]\nonumber\\
&&-\Bigl(1+{2\alpha_s(2t)\over\pi}\Bigr)\langle\bar qq\rangle_{2t}
+{\langle\bar qg_s \sigma\cdot Gq\rangle_{2t}\over 16 t^2}
\biggr\} \,,
\end{eqnarray}
which can be obtained from the two-point sum rule 
approach~\cite{BB,bbbd,Neubert2}.
For the numerical analysis,
we use the following values of parameters~\cite{yang1,yang2}
\beq
&&\langle \bar qq\rangle_{\mu=1~{\rm GeV}}=-(240~{\rm MeV})^3\,, \nonumber\\
&&\langle \alpha_s G^2 \rangle_{\mu=1~{\rm GeV}} =0.0377~{\rm GeV^4} 
\,,\nonumber\\
&&\langle \bar qg_s\sigma_{\mu\nu} G^{\mu\nu} q\rangle_{\mu=1~{\rm GeV}}=
(0.8~{\rm GeV^2})\times \langle \bar qq\rangle_{\mu=1~{\rm GeV}}  \,,
\eeq
as input and neglect the finite 
part of radiative one loop corrections. Since 
in our convention $D_\mu=\partial_\mu-ig_s A_\mu$, we have $\langle g_s 
\bar q\sigma\cdot Gq\rangle= m^2_0\langle\bar qq\rangle$.
Next, in order to determine the thresholds
$\omega_{i,j}$ we employ the $B$ meson decay
constant $f_B=(185\pm 25\pm 17)~{\rm MeV}$ obtained from a recent lattice-QCD
calculation~\cite{lattice} and the relation~\cite{Ball}
\beq
f_B ={F(m_b)\over \sqrt {m_B}}\Bigl( 1-{2\over 3}
{\alpha_s(m_b)\over \pi }\Bigr) \Bigl(1-{(0.8\sim 1.1 )~{\rm GeV}\over m_b}
\Bigr)\,,
\eeq
that takes into account QCD and $1/m_b$ corrections.
Using the relation between $F(m_b)$ and $F(\mu)$ given by
Eq.~(\ref{RGevo}) and $m_b=(4.85\pm 0.25)$~GeV, 
we obtain
\beq \label{Fresult}
F(\mu=1~{\rm GeV})
\cong (0.34\sim 0.52)~{\rm GeV^{3/2}} \,.
\eeq
Since the $\bar\Lambda$ parameter in Eq.~(\ref{F}) can be 
replaced by the $\bar\Lambda$ sum rule obtained by
applying the differential operator $t^2 \partial \ln/\partial t$
to both sides of Eq.~(\ref{F}), the $F(\mu)$ sum rule can
be rewritten as
\beq  \label{newF}
F^2(\mu)={\rm (right\ hand\ side\ of\ Eq.~(\ref{F}))}
\times {\rm exp} [t\, {\partial\over \partial t} {\rm ln}{\rm (right\ hand\
side\ of\ Eq.~(\ref{F})) }]\,,
\eeq
which is $\bar\Lambda$-free.
Then using the result (\ref{Fresult}) as  input,
the threshold $\omega_0$ in the $F(\mu)$ sum rule,
Eq.~(\ref{newF}), is determined.
The result for $\omega_0$ is  $1.25-1.65~{\rm GeV}$. A larger 
$F(\mu=1~\rm{GeV})$ corresponds to a larger $\omega_0$.
The working Borel window lies in the region
$0.6~{\rm GeV} <t < 1~{\rm GeV}$, which turns out to be a reasonable
choice~\cite{bbbd}.
Substituting the value of $\omega_0$ back into the $\bar\Lambda$ sum rule,
we obtain
 $\bar \Lambda=0.48-0.76~{\rm GeV}$ in the Borel window 
$0.6~{\rm GeV} <t < 1~{\rm GeV}$. This result is consistent with
the choice $m_b=(4.85\pm 0.25)$~GeV, recalling that in the heavy 
quark limit, $\bar\Lambda=m_B-m_b$.
To extract the $d_j^{v(i)}$ sum rules, one can take the ratio of 
Eq.~(\ref{F}) and Eq.~(\ref{rule1}) to eliminate the contribution
of $F^2/ {\rm exp}(\bar \Lambda /t)$.
This means one has chosen the same $\bar \Lambda$
both in Eq.~(\ref{F}) and Eq.~(\ref{rule1}).
Since quark-hadron duality is the basic assumption in the QCD sum rule
approach, we expect that the same result of $\bar \Lambda$ also
can be obtained using the $\bar\Lambda$ sum rules derived
from Eq.~(\ref{rule1}) (see \cite{BL,yang1} for a further discussion).
This property can help us to determine consistently the threshold in
3-point sum rule, Eq.~(\ref{rule1}).
Therefore, we can apply the differential operator
$t^2 \partial \ln/\partial t$
to both sides of Eq.~(\ref{rule1}), the $d^{v(i)}$ sum rule, to
obtain new $\bar \Lambda$ sum rules. The requirement of producing
a reasonable value for $\bar \Lambda$, say $0.48-0.76~{\rm GeV}$, provides 
severe constraints
on the choices of $\omega_{i,j}$. With a careful study, we find that
the best choice in our analysis is
\beq \label{omega3pt}
\omega_{i,1}=-0.02~{\rm GeV} +\omega_0\,, \quad
\omega_{1,2}=-0.5~{\rm GeV}+\omega_0 \,, \quad
\omega_{2,2}=-0.22~{\rm GeV}+\omega_0 \,. 
\eeq
Applying the above relations with $\omega_0=(1.25 \sim 1.65)~{\rm GeV}$ and
substituting $F(\mu)$ in Eq.~(\ref{rule1}) by (\ref{F}), we study numerically
the $d_j^{v(i)}$ sum rules. In Fig.~2,
we plot $B_i^v$ and $\varepsilon_i^v$ as a function $t$, 
where $B_i^v=8d_1^{v(i)}/9 + 2d_2^{v(i)}/3,\ $ and
$\varepsilon_i^v=-4d_1^{v(i)}/27+ 8d_2^{v(i)}/9$.
The dashed and solid curves stand for $B_i^{v}$ and
$\varepsilon_i^{v}$, respectively, where we have used
$\omega_0=1.4~ {\rm GeV}$ (the corresponding
decay constant is $f_B=175\sim 195~ {\rm MeV}$
or $F(\mu)=0.405\pm 0.005~ {\rm GeV}^{3/2}$).
The final results for the hadronic parameters $B_i$ and $\varepsilon_i$
are (see Fig.~2) 
\footnote{For comparison, the sum rule results obtained in \cite{BLLS} and 
\cite{Chern}
are $\varepsilon_1^{\rm BLLS}(\mu)=-0.041\pm 0.022$, $\varepsilon_2^{\rm BLLS}
(\mu)=0.061\pm 
0.035$ \cite{BLLS} and $\varepsilon^{\rm C}_1(\mu)\approx -0.15$, 
$\varepsilon^{\rm C}_2(\mu)\approx 0$ \cite{Chern}, respectively, with $\mu$ 
being a typical hadronic scale $\sim 0.70$ GeV. Note that the definition of
$B_i(\mu)$ and $\varepsilon_i(\mu)$ in \cite{BLLS,Chern} is different from 
ours by a factor of $F^2(m_b)/F^2(\mu)$, that is, $\varepsilon_i^{\rm 
BLLS,C}(\mu)=\varepsilon_i(\mu)\times F^2(m_b)/F^2(\mu)$.}
\beq
 B_1^{v}(\mu=1~{\rm GeV})=0.60\pm 0.02, \qquad &&
B_2^{v}(\mu=1~{\rm GeV})=0.61\pm 0.01, \nonumber \\
 \varepsilon_1^{v}(\mu=1~{\rm GeV})=-0.08\pm 0.01,   \qquad  &&
\varepsilon_2^{v}(\mu=1~{\rm GeV})=-0.024\pm 0.006.
\eeq
The numerical errors come mainly from the uncertainty of 
$\omega_0=1.25\sim 1.65$~GeV.
Some intrinsic errors of the sum rule approach, say quark-hadron
duality or $\alpha_s$ corrections, will not be considered here.

Substituting the above results into Eq.~(\ref{lowpara}) yields
\beq   \label{biei}
B_1(m_b)=0.96\pm 0.04 + { O}(1/m_b)\,, && \qquad
B_2(m_b)=0.95\pm 0.02 + { O}(1/m_b)\,, \nonumber\\
\varepsilon_1(m_b)=-0.14\pm 0.01 +{ O}(1/m_b)\,,  && \qquad
\varepsilon_2(m_b)=-0.08\pm 0.01 +{ O}(1/m_b)\,.
\eeq
It follows from Eq.~(\ref{ratios}) that
\beq
&& \frac{\tau (B^-)}{\tau(B_d)} = 1.11 \pm 0.02\,,   \nonumber \\
&& \frac{\tau (B_s)}{\tau(B_d)}\approx 1\,,  \nonumber\\
&& \frac{\tau (\Lambda_b)}{\tau (B_d)} = 0.99 -
\Big(\frac{f_B}{185~{\rm MeV}}\Bigr)^2
(0.007+0.020\, \tilde B)\, r \,,
\eeq
to the order of $1/m_b^3$. Note that we have neglected the corrections of 
SU(3) symmetry breaking to the nonspectator effects in $\tau (B_s)/\tau(B_d)$.
We see that the prediction for $\tau(B^-)/\tau
(B_d)$ is in agreement with the current world average:
$\tau(B^-)/\tau(B_d)$~=1.07$\pm$ 0.04 \cite{LEP}, whereas the
heavy-quark-expansion-based result for $\tau(B_s)/\tau(B_d)$ deviates
somewhat from the central value of the world average
\footnote{For example, the neutral $B$ meson lifetimes are measured at CDF 
to be \cite{Abe}: $\tau(B_d)=1.58\pm 0.09\pm 0.02$ ps and $\tau(B_s)= 1.34
^{+0.23}_{-0.19}\pm 0.05$ ps.}: $0.95\pm 0.05$.
Thus it is urgent to carry out more precise measurements of the $B_s$ lifetime.
Using the existing sum rule estimate for the parameter $r$ \cite{Col} together
with $\tilde B=1$ gives $\tau(\Lambda_b)/\tau(B_d)\geq 0.98$. Therefore,
the $1/m_b^3$ nonspectator corrections are not responsible for the
observed lifetime difference between the $\Lambda_b$ and $B_d$.

\section{Discussions and Conclusions}
The prediction of $B$ meson lifetime ratios depends on the nonspectator 
effects of order $16\pi^2/m_b^3$
in the heavy quark expansion. These effects can be parametrized in 
terms of the hadronic parameters $B_1$, $B_2$, 
$\varepsilon_1$ and $\varepsilon_2$, where $B_1$ and $B_2$ characterize
the matrix elements of color singlet-singlet four-quark operators and 
$\varepsilon_1$ and $\varepsilon_2$ the matrix elements of color 
octet-octet operators.

  As emphasized in \cite{Cheng}, one should not be contented with the 
agreement between theory and experiment for the lifetime ratio $\tau(B^-)/
\tau(B_d)$. In order to test the OPE approach for inclusive nonleptonic
decay, it is even more important to calculate the absolute decay
widths of the $B$ mesons and compare them with the data. From (\ref{num})
 and (\ref{biei}) and considering the contributions of the nonspectator
effects, we obtain
\beq \label{width}
\Gamma_{\rm tot}(B_d) &=& \,(3.61^{+1.04}_{-0.84})
\times 10^{-13}\,{\rm GeV}, \nonumber\\  
\Gamma_{\rm tot}(B^-) &=& \,(3.34^{+1.04}_{-0.84})
\times 10^{-13}\,{\rm GeV},  
\eeq
noting that the next-to-leading QCD radiative
correction to the inclusive decay width has been included.
The absolute decay widths strongly depend on the value of the $b$ quark
mass.

 The problem with the absolute
decay width $\Gamma(B)$ is intimately related to the $B$ meson semileptonic
branching ratio ${\cal B}_{\rm SL}$. 
Unlike the semileptonic decays, the heavy quark expansion in inclusive
nonleptonic decay is {\it a priori} not justified due to the absence of an 
analytic continuation into the complex plane and hence local 
duality has to be invoked in
order to apply the OPE directly in the physical region. If the shorter
lifetime of the $\Lambda_b$ relative to that of the $B_d$ meson is confirmed
in the future and/or if the lifetime ratio $\tau(B_s)/\tau(B_d)$ is observed 
to be approximately unity, then it is very likely that local quark-hadron 
duality is violated in nonleptonic decays. It should be stressed that local 
duality is {\it exact} in the heavy quark limit, but its systematic $1/m_Q$ 
expansion is still lacking. Empirically, it has been suggested in 
\cite{Altarelli} that the presence of linear $1/m_b$ correction, described
by the ansatz that the $b$ quark mass $m_b$ is replaced by the decaying
bottom hadron mass $m_{H_b}$ in the $m_b^5$ factor in front of all
nonleptonic widths, will account for the observed lifetime difference
between the $\Lambda_b$ and $B_d$. To be specific, the ansatz $\Gamma_{\rm NL}
\to\Gamma_{\rm NL}(m_{H_b}/m_b)^5$ will lead to the results \cite{Cheng}:
\beq
{\tau(\Lambda_b)\over \tau(B_d)} = 0.76\,,  \qquad \quad
{\tau(B_s)\over \tau(B_d)} = 0.94\,.
\eeq
This simple prescription not only solves the lifetime ratio problem but also
provides the correct absolute decay widths for the $\Lambda_b$ and the $B$
mesons. The predicted lifetime hierarchy 
\beq
\tau(\Lambda_b)>\tau(\Xi^-_b)>\tau(\Xi^0_b)>\tau(\Omega_b)
\eeq
for bottom baryons is in sharp contrast to the OPE-based lifetime pattern
\cite{Cheng}:
\beq
\tau(\Omega_b)\simeq \tau(\Xi^-_b)>\tau(\Lambda_b)\simeq\tau(\Xi_b^0).
\eeq
Of course, whether this empirical ansatz truly works or whether it
can be justified in a more fundamental way (see, for example, \cite{Jin})
remains to be investigated.
Nevertheless, it is worth emphasizing that, although a linear $1/m_Q$ 
correction to the inclusive nonleptonic decay rate is possible 
\cite{Colangelo,Grinstein},  the violation of
local quark-hadron duality does not necessarily imply the presence of
$1/m_Q$ terms in inclusive widths and hence the aforementioned ansatz.

To conclude, we have derived in heavy quark effective theory
the renormalization-group improved sum rules for the hadronic parameters 
$B_1$, $B_2$, $\varepsilon_1$, and $\varepsilon_2$ appearing in the
matrix element of four-quark operators. 
The results are $B_1(m_b)=0.96\pm 0.04$, $B_2(m_b)=0.95\pm 0.02$,
$\varepsilon_1(m_b)=-0.14\pm 0.01$ and $\varepsilon_2(m_b)=-0.08\pm 0.01$
to the zeroth order in $1/m_b$.
The resultant $B$-meson lifetime ratios are $\tau(B^-)/\tau(B_d)=1.11\pm 
0.02$ and $\tau(B_s)/\tau(B_d)\approx 1$.

\acknowledgments
This work was supported in part by the National Science Council of
R.O.C. under Grant No. NSC87-2112-M-001-048.


\newpage

\noindent
\begin{figure}
\vspace{4cm}
\epsfbox{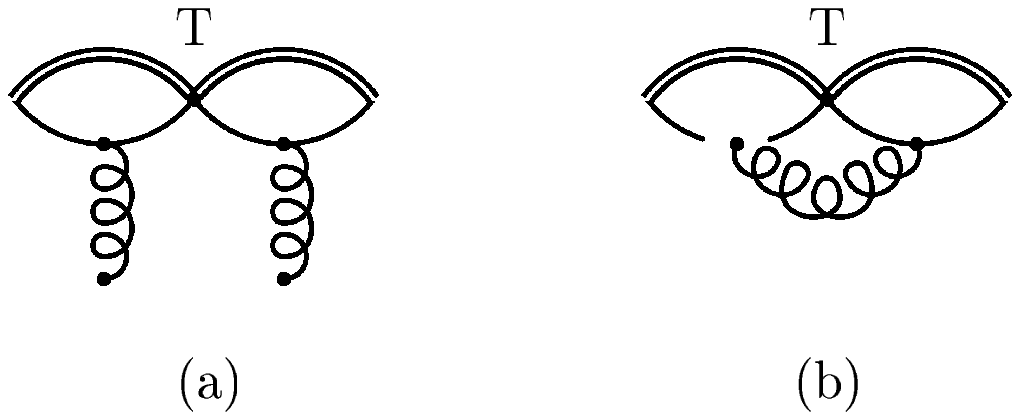}
\vskip 1cm 
\caption{
The main diagrams contributing to 
${\rm OPE}_{\varepsilon_i}$ [cf. Eq.~(3.24)]:
(a) the contribution from the gluon condensate, and
(b) the contribution from the quark-gluon mixed condensate.
In (b) the mirror-symmetric diagram is included in the calculation.
The double lines denote heavy quarks in HQET.}
\end{figure}

\noindent
\begin{figure}
\epsfbox{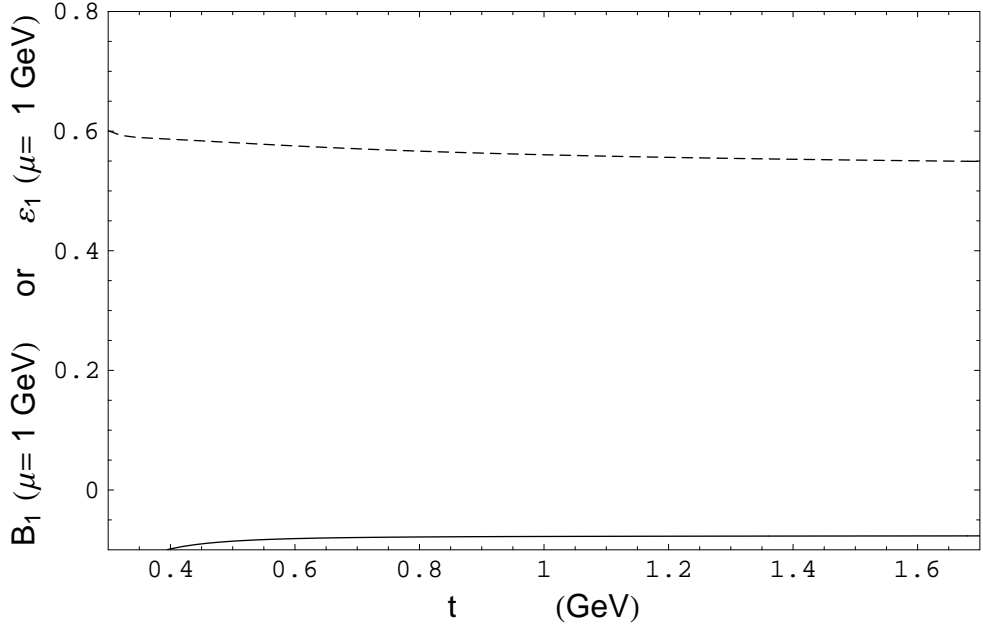}
\vspace{10mm}
\epsfbox{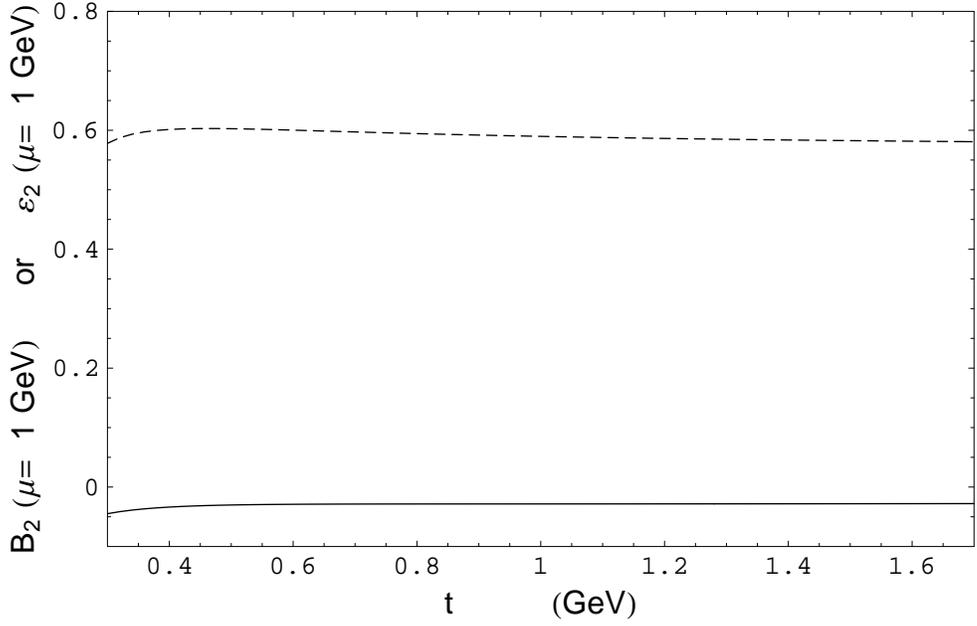}
\vspace{10mm}
\caption{
$B_i^v(\mu)$ and $\varepsilon_i^v(\mu)$ as a function $t$, 
where $B_i^v=8d_1^{v(i)}/9 + 2d_2^{v(i)}/3,\ $ and
$\varepsilon_i^v=-4d_1^{v(i)}/27+ 8d_2^{v(i)}/9$.
The dashed and solid curves stand for $B_i^{v}$ and
$\varepsilon_i^{v}$, respectively. Here we have used
$\omega_0=1.4~ {\rm GeV}$ and Eq.~(3.33).}
\end{figure}
\end{document}